\documentstyle[prl,aps,preprint]{revtex}
\newcommand{\dd}{{d}}
\newcommand{\tr}{{\rm tr}}
\newcommand{\Tr}{{\rm Tr}}

\newcommand{\thru}[1]{\mathrel{\mathop{#1\!\!\!/}}}

\newcommand{\bfD}{\mbox{\boldmath $D$}}
\newcommand{\bfDt}{{\tilde{\bfD}}}

\newcommand{\bfJ}{\mbox{\boldmath $J$}}
\newcommand{\bfJt}{{\tilde{\bfJ}}}
\newcommand{\bfM}{\mbox{\boldmath $M$}}
\newcommand{\bfMt}{{\tilde{\bfM}}}
\newcommand{\bfN}{\mbox{\boldmath $N$}}
\newcommand{\bfQ}{\mbox{\boldmath $Q$}}
\newcommand{\bfQt}{{\tilde{\bfQ}}}

\newcommand{\bfY}{\mbox{\boldmath $Y$}}

\newcommand{\Ac}{{\cal A}}

\begin{document}

\draft
\tighten
\def\footnoterule{\kern-3pt \hrule width\hsize \kern3pt}

\title{
Chiral and scale anomalies of non local Dirac operators
}

\author{E. Ruiz Arriola and L.L. Salcedo}

\address{
{~} \\
Departamento de F\'{\i}sica Moderna \\
Universidad de Granada \\
E-18071 Granada, Spain
}

\date{\today}
\maketitle

\thispagestyle{empty}

\begin{abstract}
The chiral and scale anomalies of a very general class of non local
Dirac operators are computed using the $\zeta$-function definition of
the fermionic determinant. For the axial anomaly all new terms
introduced by the non locality are shown to be removable by counter
terms and such counter terms are also explicitly computed. It is
verified that the non local Dirac operators have the standard minimal
anomaly in Bardeen's form.
\end{abstract}


\pacs{PACS numbers:\ \ 11.10.Lm, 11.30.Rd, 11.15.-q, 12.39.Fe}

\section{Introduction }\label{sec:0}

Local field theories provide the traditional setup  where the
implementation of space-time symmetries becomes rather simple. On the
other hand, effective theories are not necessarily local, although an
appropriate choice of degrees of freedom can make them local
\cite{We79}. An outstanding example is QCD in the domain of low
hadronic energies, where light quarks and gluons are dressed by the
interaction making the effective theory look highly nonlocal
\cite{MP78} in terms of these degrees of freedom and hence a sort of
dynamical perturbation theory would be
needed \cite{PS79}. This has produced a wealth of work mainly based on
Dyson-Schwinger equations  properly constrained by the
relevant Ward and Slavnov-Taylor identities \cite{RW94}.
In this respect anomalies provide an interesting playground to study the
interplay between low and high energies both in the local and in the
nonlocal case. They are triggered by the ultraviolet regulators which
unavoidably violate some classical symmetries but their physical
effect is formulated as a low energy theorem. The question
whether or not the nonlocal interaction can be implemented without
spoiling the anomaly has been previously discussed \cite{RC88,HT89,BR93}
for some specific processes like e.g. $\pi^0 \to 2 \gamma $, $ \gamma
\to 3 \pi $ and $ 2 K \to 3 \pi $ and regarding the chiral anomaly. In
this paper we study the question for all processes and both chiral and
scale anomalies. Rather than computing specific processes one by one we
just prove that the new terms generated by the non locality can be
subtracted by adding suitable counterterms.

\section{Non local Dirac operators}\label{sec:1}

We will consider Dirac fermions in the flat $D$-dimensional Euclidean
space-time R$^D$ endowed with internal degrees of freedom collectively
referred to as ``flavor''. The class of Dirac operators to be
considered here is
\begin{equation}
\bfD=\bfD_L + \bfM \,.
\end{equation}
The term $\bfD_L$, the local component of $\bfD$, is a standard Dirac
operator
\begin{equation}
\bfD_L=  \gamma_\mu P_\mu+\bfY
\end{equation}
The Dirac gamma matrices are anti-Hermitian (we will follow the
conventions of \cite{Sa96}), $P_\mu=i\partial_\mu$ and $\bfY$ is an
arbitrary matrix-valued function in flavor and Dirac spaces. It will
be convenient to regard $\bfY$ as a function of the position operators
$X_\mu$, defined by $X_\mu\psi(x)=x_\mu\psi(x)$, so that $\bfY$ is a
multiplicative operator in the Hilbert space of fermions.
\begin{equation}
(\bfY\psi)(x)=\bfY(x)\psi(x)\,.
\end{equation}
The term $\bfM$ is a purely non local, more precisely bilocal, operator
also with arbitrary structure in flavor and Dirac spaces,
\begin{equation}
(\bfM\psi)(x)=\int \dd^Dy\bfM(x,y)\psi(y)\,.
\label{eq:4}
\end{equation}
By purely non local we mean that $\bfM$ is softer in the ultraviolet
sector than any multiplicative operator, that is, the distribution
$\bfM(x,y)$ is less singular than the Dirac delta $\delta(x-y)$. More
restrictive assumptions on $\bfM$ will be made below. Further
restrictions on the form of $\bfY$ and $\bfM$ come from imposing
hermiticity of the associated Hamiltonian in Minkowski space (that is,
the hermiticity of $\gamma^0\bfD$). In even dimensions, the Euclidean
Dirac operator can be split into the components with and without
$\gamma_5$, $\bfD_-$ and $\bfD_+$ respectively, then unitarity
requires $\bfD_\pm^\dagger=\pm\gamma_5\bfD_\pm\gamma_5$. In the odd
dimensional case, $\bfD_+$ and $\bfD_-$ corresponds to an even or odd
number of Dirac matrices, respectively, and the unitarity condition
becomes $\bfD_\pm^\dagger=\pm\bfD_\pm$.

Many of the concepts used for standard local Dirac operators apply
directly to the non local case. We define a symmetry as any
transformation of $\psi(x)$ and $\bar\psi(x)$ that can be compensated
by a corresponding transformations of the external fields $\bfY$ and
$\bfM$ (within the class of fields considered) so that the action
$\int \dd^Dx\bar\psi\bfD\psi$ remains invariant. Presently, we will
consider chiral transformations in two and four dimensions. Scale
transformations will be treated in section~\ref{sec:4}. Chiral
transformations are defined as in the local case, namely,
\begin{equation}
\bfD \to e^{i\beta-i\alpha\gamma_5}\bfD e^{-i\beta-i\alpha\gamma_5}\,,
\end{equation}
where $\alpha(x)$ and $\beta(x)$ are Hermitian matrices in flavor
space only, regarded as multiplicative operators on the fermionic
wave functions. The particular cases $\alpha=0$ and $\beta=0$
correspond to vector and axial transformations, respectively. In the
infinitesimal case
\begin{equation}
\delta\bfD = \delta_V\bfD+ \delta_A\bfD=
[i\beta,\bfD]-\{i\alpha\gamma_5,\bfD\}\,.
\end{equation}
Because the chiral transformations are local, both $\bfD_L$ and $\bfM$
transform covariantly separately, that is,
\begin{equation}
\delta\bfD_L = [i\beta,\bfD_L]-\{i\alpha\gamma_5,\bfD_L\}\,, \quad
\delta\bfM = [i\beta,\bfM]-\{i\alpha\gamma_5,\bfM\}\,.
\end{equation}
Note that the bilocal structure of $M$ implies that local factors at
each side of the operator are taken at different points, i.e.
$ \bfM(x,x') \to \
         e^{i\beta(x)-i\alpha(x)\gamma_5} \bfM(x,x')
e^{-i\beta(x')-i\alpha(x') \gamma_5}$.
The effective action of the fermions in presence of the external
fields $\bfY$ and $\bfM$ is defined as in the local case, namely
\begin{equation}
W(\bfD)=
-\log\int {\cal D}\bar\psi{\cal D}\psi \exp \left\{
-\int \dd^Dx\,\bar\psi(x)\bfD\psi(x)\right\}
=-\Tr\log\bfD\,.
\end{equation}
Here, $\Tr$ stands for trace over all degrees of freedom and some
renormalization of the ultraviolet divergences is understood. The
(consistent) anomaly is defined as the variation of the effective
action under infinitesimal chiral transformations. Since we will be
considering a $\zeta$-function renormalization of $W$, there will be
no vector anomaly,
\begin{equation}
\delta_VW=0, \quad \delta_AW=\Ac_A\,.
\end{equation}
Correspondingly, the same current conservation formulas valid for the
local case can be written here,
\begin{eqnarray}
0 &=& \int\dd^Dx\,\langle\bar\psi(x)[i\beta,\bfD]\psi(x)\rangle_Q
\\
-\Ac_A &=&
\int\dd^Dx\,\langle\bar\psi(x)\{i\alpha\gamma_5,\bfD\}\psi(x)\rangle_Q \,.
\end{eqnarray}
(The symbol $\langle\ \rangle_Q$ stands for quantum vacuum expectation
value.) In particular the term $\gamma_\mu P_\mu$ in $\bfD$ in the
right-hand side yields, after integration by parts, the divergence of
the fermionic vector and axial currents whereas the other terms in
$\bfD$, local and non local, represent the explicit chiral symmetry
breaking due to the external fields. On the other hand, the left-hand
side shows the anomalous breaking of the axial current
conservation. As will be shown below, the effective action can be
renormalized so that only the local fields $\bfY$ contribute to the
anomaly, and moreover, only the standard minimal Bardeen's axial
anomaly needs to be retained.

For the purpose of doing detailed calculations we will assume that the
non local operator $\bfM$ admits an expansion in inverse powers of
$P_\mu$ for large $P_\mu$ of the form
\begin{equation}
\bfM=\bfM_\mu\frac{P_\mu}{P^2}+
\bfM_{\mu\nu}\frac{P_\mu P_\nu}{P^4}+
\bfM_{\mu\nu\rho}\frac{P_\mu P_\nu P_\rho}{P^6}+
\cdots
\label{eq:12}
\end{equation}
The coefficients $\bfM_{\mu_1\dots\mu_n}$ are multiplicative operators
and they are completely symmetric under permutation of indices. For
convenience the $P_\mu$ has been put at the right. It should be noted
that this choice does not exhaust all possible non local
operators. For instance, for each given $k=0,1,2,\dots$, the class of
operators
\begin{equation}
\bfM^{(k)}=\sum_n\sum_{\mu,\alpha}
\bfM_{\mu_1\dots\mu_n}^{\alpha_1\dots\alpha_{2k}}
\frac{P_{\mu_1}\cdots P_{\mu_n}}{P^{2n}}
\frac{P_{\alpha_1}\cdots P_{\alpha_{2k}}}{P^{2k}}
\end{equation}
includes all classes with lower index $k$ as particular cases. Our
choice $k=0$ is the simplest one but it is still non trivial and
enjoys the essential property of being closed under chiral
transformations \footnote{ \footnotesize This choice is also realistic
since it accommodates the operator product
expansion estimate of the quark self-energy $  \Sigma (p^2 ) \sim_{p^2
\to \infty} ( {\rm  log} p^2 )^{d-1} / p^2 $ with $d$ the anomalous
dimension of the quark
 condensate $\bar \psi \psi $ (see ref.\cite{Po76})}.   The
better way to obtain the transformation
properties of $\bfM$ is by introducing a family of operators
associated to $\bfM$ as
\begin{equation}
\bfMt(p)= e^{ipX}\bfM e^{-ipX}
\end{equation}
where the momentum $p_\mu$ is just a constant c-number. Effectively,
$\bfMt(p)$ corresponds to make the replacement $P_\mu\to P_\mu+p_\mu$
in $\bfM$. The function $\bfMt(p)$ admits an expansion in inverse
powers of $p_\mu$ similar to that in eq.~(\ref{eq:12}), namely
\begin{equation}
\bfMt(p)=\bfMt_\mu\frac{p_\mu}{p^2}+ \bfMt_{\mu\nu}\frac{p_\mu
p_\nu}{p^4}+ \cdots\,.
\label{eq:15}
\end{equation}
The two lowest coefficients are given by
\begin{eqnarray}
\bfMt_\mu &=& \bfM_\mu \,, \nonumber \\
\bfMt_{\mu\nu} &=&
\bfM_{\mu\nu}+t_{\mu\nu\rho\sigma}\bfM_\rho P_\sigma \,,
\end{eqnarray}
where we have introduced $t_{\mu\nu\rho\sigma}=
\delta_{\mu\nu}\delta_{\rho\sigma} -
\delta_{\mu\rho}\delta_{\nu\sigma}-\delta_{\mu\sigma}\delta_{\nu\rho}$.
It should be noted that the coefficients $\bfMt_{\mu_1\dots\mu_n}$ are
not multiplicative operators. One useful property of $\bfMt(p)$ is
that it transforms covariantly under chiral transformations. Indeed,
if $\bfM_\Omega=\Omega_1\bfM\Omega_2$ for two multiplicative operators
$\Omega_{1,2}$,
\begin{equation}
\bfMt_\Omega(p)= e^{ipX}\bfM_\Omega e^{-ipX}
=\Omega_1\bfMt(p)\Omega_2\,.
\end{equation}
As a consequence, the coefficients are also chiral covariant
\begin{equation}
\delta\bfMt_{\mu_1\dots\mu_n} = [i\beta,\bfMt_{\mu_1\dots\mu_n}]
-\{i\alpha\gamma_5,\bfMt_{\mu_1\dots\mu_n}\}\,.
\end{equation}
From here it is immediate to derive the transformation of the
original coefficients $\bfM_{\mu_1\dots\mu_n}$. For the
two lowest order coefficients one finds
\begin{eqnarray}
\delta\bfM_\mu &=&
[i\beta,\bfM_\mu]-\{i\alpha\gamma_5,\bfM_\mu\}\,, \\
\delta\bfM_{\mu\nu} &=&
[i\beta,\bfM_{\mu\nu}]-\{i\alpha\gamma_5,\bfM_{\mu\nu}\}
+t_{\mu\nu\rho\sigma}
\bfM_\rho(\partial_\sigma\beta+\partial_\sigma\alpha\gamma_5) \,. \nonumber
\end{eqnarray}
In general, the variation of each coefficients involves those of lower
order. This shows that the class of non local operators considered
carries a representation of the chiral group.

There is another essential requirement which is also satisfied by the
particular class of non local operators $\bfM$ considered, namely, it
is closed under Hermitian conjugation (see section~\ref{sec:3}) and so
the coefficients can be chosen so as to satisfy the unitarity
requirement stated above. Another remark is that the classes of
operators corresponding to set to zero the first $n$ coefficients in
$\bfM$ are also closed under chiral transformations and Hermitian
conjugation.

\section{The Axial anomaly}\label{sec:2}

In order to compute the axial anomaly, we will adopt the
$\zeta$-function renormalization prescription combined with an
asymmetric Wigner transformation. This method, as well as
several of its applications, is presented in great detail in
\cite{Sa96}. Since the techniques required in the present non local
case are an immediate extension of those used in that reference here
we will emphasize only the new issues introduced by the non
locality. The $\zeta$-function effective action is given
by~\cite{Se67,Ha77}
\begin{equation}
W(\bfD) = -\frac{\dd}{\dd s} \Tr\left(\bfD^s\right)_{s=0}\,,
\end{equation}
where $s=0$ is to be understood as an analytical extension on $s$ from
the ultraviolet convergent region ${\rm Re}(s)<-D$. The key point is
that for sufficiently negative $s$ there are no ultraviolet
divergences and formal operations become justified. By construction, the
$\zeta$-function renormalized effective action is invariant under all
symmetry transformations associated to similarity transformations of
$\bfD$, thus in particular it is vector gauge invariant. On the other
hand the axial anomaly takes the form
\begin{equation}
\Ac_A = \Tr\left(2i\alpha\gamma_5\bfD^s\right)_{s=0}\,.
\end{equation}
The operator $\bfD^s$ can be obtained from
\begin{equation}
\bfD^s = -\int_\Gamma\frac{\dd z}{2\pi i} \frac{z^s}{\bfD-z}
\end{equation}
where the integration path $\Gamma$ starts at $-\infty$, follows the
real negative axis, encircles the origin $z=0$ clockwise and goes back
to $-\infty$. Using the Wigner transformation technique~\cite{Sa96},
the anomaly can be written as (a similar expression holds for the
effective action)
\begin{equation}
\Ac_A = -\int\frac{\dd^Dp}{(2\pi)^D}\int_\Gamma\frac{\dd z}{2\pi i}z^s
\tr\langle 0|2i\alpha\gamma_5\frac{1}{\bfDt(p)-z}|0\rangle\Big|_{s=0}\,.
\end{equation}
Here $\tr$ stands for trace over Dirac and flavor degrees of freedom,
$|0\rangle$ is the zero momentum state normalized as $\langle
x|0\rangle=1$, thus $P_\mu|0\rangle=\langle0|P_\mu=0$. Further
\begin{equation}
\bfDt(p)= e^{ipX}\bfD e^{-ipX} = \thru{p}+\bfD_L+\bfMt(p)\,.
\end{equation}
The integration over $z$ should be performed first, since it defines
the operator $\bfD^s$, then the integral over $p$ which corresponds to
take the trace over space-time degrees of freedom and finally $s$ is
to be analytically extended to $s=0$. The simplest way to proceed is
to introduce a mass term, i.e., to apply the formula to the Dirac
operator $\bfD+m$ and then make an expansion in powers of
$\bfD_L+\bfMt(p)$, letting $m\to 0$ at the end. In this way the
following expression is derived
\begin{equation}
\Ac_A = \sum_{N \ge 0}
\int\frac{\dd^Dp}{(2\pi)^D}
\int_\Gamma\frac{\dd z}{2\pi i}z^s
\tr\langle 0|2i\alpha\gamma_5
\frac{(\bfD_L+\bfMt(p))\left((\thru{p}+z-m)(\bfD_L+\bfMt(p))\right)^N}
{(p^2+(z-m)^2)^{N+1}}
|0\rangle\Big|_{s=0, m=0}\,.
\end{equation}
This formula has been simplified using the cyclic property for the
trace in Dirac space.

From the expression it is clear that for sufficiently large $N$ the
integrals become ultraviolet convergent. When this happens $s$ can be
set to zero directly and $\Gamma$ no longer encloses any singularity
thus the integral vanishes. Using this insight we can expand
$\bfMt(p)$ in inverse powers of $p_\mu$ and keep only the divergent
terms. Using eq.~(\ref{eq:15}), the anomaly can be written as a sum of
monomials each given as a product of $\bfD_L$ and the various
coefficients $\bfMt_{\mu_1\dots\mu_n}$ raised to different
powers. The canonical dimension of the monomial, $g$, is obtained
noting that the canonical dimension of $\bfD_L$ is 1 and that of
$\bfM_{\mu_1\dots\mu_n}$ is $n+1$ (the various factors of $m$, $z$ and
$p_\mu$ do not count to compute $g$). Doing the usual dimensional
analysis, it is easily established that after integration each
monomial comes with a factor $m^\gamma$ where $\gamma$ is the degree
of divergence of the term and further $\gamma=D-g+s$. As noted above,
terms with a negative degree of divergence vanish. On the other hand,
the terms with a positive degree of divergence also vanish after
taking $m=0$, since $m$ is raised to a positive power. In summary, the
only contributions which need to be retained are those logarithmically
divergent which correspond to monomials with scale dimension $g=D$. In
two dimensions, the only relevant terms are those of the form
$\bfD_L^2+\bfMt_\mu$, whereas in four dimensions they are $\bfD_L^4+
\bfD_L^2\bfMt_\mu+\bfD_L\bfMt_{\mu\nu}+\bfMt_\mu^2+
\bfMt_{\mu\nu\rho}$. A further restriction comes from Euclidean
rotational invariance on $p_\mu$. As a consequence, it is immediate to
see that the terms $\bfMt_\mu$ in two dimensions and
$\bfMt_{\mu\nu\rho}$ in four dimensions cancel.

After an angular average over $p_\mu$, the indicated integrals on
$p_\mu$ and $z$ can be carried out directly with the integral $I_1$
given in \cite{Sa96}. The result for the two dimensional anomaly is
\begin{equation}
\Ac_A = \langle 2i\alpha\gamma_5\bfD_L^2\rangle,
\label{eq:26}
\end{equation}
and in four dimensions
\begin{eqnarray}
{\cal A}_A &=& \Big\langle 2i\alpha\gamma_5 \Big(
\frac{1}{2}\bfD_L^4 + \frac{1}{12}\bfD_L\{\gamma_\mu,\bfD_L\}^2\bfD_L
\nonumber\\ &&
+\frac{1}{8}\left(\bfMt_\mu\{\gamma_\mu,\bfD_L\}\bfD_L
+\bfD_L\{\gamma_\mu,\bfMt_\mu\}\bfD_L
+\bfD_L\{\gamma_\mu,\bfD_L\}\bfMt_\mu \right)
\nonumber\\ &&
+\frac{1}{4}\bfMt_\mu^2 +\frac{1}{4}\{\bfD_L,\bfMt_{\mu\mu}\}
\Big) \Big\rangle
\label{eq:27}
\end{eqnarray}
The notation $\langle f\rangle$ stands for
\begin{equation}
\langle f\rangle = \frac{1}{(4\pi)^{D/2}}\tr\langle 0|f(X)|0\rangle \,.
\end{equation}

Observe that, even for non local Dirac operators, the anomaly is a
local polynomial of dimension $D$ constructed with $P_\mu$ and the
external fields $\bfY$ and $\bfM_{\mu_1\dots\mu_n}$. This is a
general property of all anomalies since only ultraviolet divergent
terms can contribute to them. This puts a restriction to the number of
non local coefficients that can appear in the axial anomaly, namely,
only coefficients of dimension at most $D$ can be relevant. In terms
of the kernel $\bfM(x,y)$ in eq.~(\ref{eq:4}), it implies that non
local components in the Dirac operator with kernels which are
piecewise continuous with jump discontinuities do not contribute to
the anomaly. The detailed calculation shows that actually the
coefficients with dimension $D$ ($\bfMt_\mu$ in two dimensions and
$\bfMt_{\mu\nu\rho}$ in four dimensions) do no have a contribution
either. In particular, in two dimensions there is no non local
contribution to the axial anomaly.

The expressions found for the anomaly can be put in a more usual form,
in terms of vector and axial fields, scalar fields, etc, by making two
observations. First, after taking the Dirac trace, the operators that
appear there are actually multiplicative, that is, all $P_\mu$ appear
inside commutators only. The simplest way to see this is by formally
replacing every $P_\mu$ by $P_\mu+a_\mu$ where $a_\mu$ is a constant
c-number, and checking that all $a_\mu$-dependence cancels. Second,
for a multiplicative operator $f(X)$,
\begin{equation}
\langle f(X)\rangle = \frac{1}{(4\pi)^{D/2}}\int d^Dx \tr f(x) \,.
\end{equation}

Since the regularization preserves vector gauge invariance, the axial
anomaly is also invariant. In our expression for the anomaly, this is
a direct consequence of the operators there being
multiplicative. Indeed, any operator $f$ constructed with the gauge
covariant blocks $\bfD_L$ and $\bfMt_{\mu_1\dots\mu_n}$ is also
covariant, i.e., $f\to \Omega f\Omega^{-1}$. If in addition $f$ is
multiplicative $\langle f\rangle$ is invariant. Note that
$\langle\,\rangle$ is not a trace and so the cyclic property does not
hold for arbitrary non multiplicative operators.

\section{Essential axial anomaly}\label{sec:3}

Presumably due to its topological connection~\cite{Al85}, the axial
anomaly is a very robust quantity. It is not affected by higher order
radiative corrections~\cite{AB69}, and remains unchanged at finite
temperature and density~\cite{Go94}. It gets no contributions from
scalar and pseudo scalar fields~\cite{Ba69}, tensor
fields~\cite{Cl83,Mi87,Sa96} or internal gauge fields, i.e,
transforming homogeneously under gauge transformations
~\cite{Bi93,Ru95}. In all known cases, the anomaly only affects the
imaginary part of the effective action and only involves vector and
axial fields. The counter terms can always be chosen so that the axial
anomaly adopts the minimal or Bardeen's form~\cite{Ba69}. Not
surprisingly, the new terms introduced in the anomaly by the non local
component of the Dirac operator are also unessential, that is, they
can be removed by adding a suitable local and polynomial counter term
to the effective action. In other words, all new terms can be derived
as the axial variation of an action which is a polynomial constructed
with the external fields $\bfY$ and $\bfM_{\mu_1\dots\mu_n}$ and their
derivatives. The canonical dimension of the polynomial can be at most
$D$.

The general proof that the anomaly can always be brought to Bardeen's
form is as follows. Let $\bfY_i(x)$ denote the various external fields which
specify the Dirac operator. Under a variation of them
\begin{equation}
\delta \bfD= \delta \bfY_i\frac{\partial\bfD}{\partial \bfY_i}\,.
\end{equation}
For convenience, we consider $\bfY_i$ as multiplicative operators and put
them at the right. The operators $\partial\bfD/\partial \bfY_i$ are
simply constants, in the sense that they do not depend on $\bfY_i$, and
are just numbers or matrices if $\bfY_i$ refers to a local degree of
freedom and contain $P_\mu$ when $\bfY_i$ is related to a non local term
of $\bfD$. Each external field defines a consistent current through
the variation of the effective action
\begin{equation}
\delta W=\int d^Dx\sum_i\tr\,(\delta \bfY_i(x)\bfJ_i(x))\,.
\end{equation}
At a formal level, the currents would be just $\bfJ_i(x)= -\langle
x|(\partial\bfD/\partial \bfY_i)\bfD^{-1}| x \rangle$. However this
expression is ultraviolet divergent. Within the $\zeta$-function
prescription the renormalized consistent currents are given by
\begin{equation}
\bfJ_i(x)= -\frac{d}{ds}\left(s\langle x|\frac{\partial\bfD}{\partial
\bfY_i}\bfD^{s-1}| x \rangle\right)_{s=0}\,.
\end{equation}
It is also possible to define the currents in a chirally covariant
manner~\cite{Ba84}. One such definition~\cite{Le85} corresponds to
take the finite part, as $\epsilon$ goes to zero, of
\begin{equation}
\bfJ_i^c(x)=
-\int_\epsilon^\infty d\lambda\langle x|\frac{\partial\bfD}{\partial \bfY_i}
\bfD^\dagger e^{-\lambda\bfD\bfD^\dagger}| x \rangle \,.
\end{equation}
(Recall that $\alpha(x)$ is Hermitian and so $\bfD^\dagger$ transform
as $\bfD^{-1}$ under axial transformations.) Due to the presence of an
essential axial anomaly, the consistent and covariant currents cannot
(all of them) coincide. If the covariant currents were consistent they
could be integrated to yield a chiral invariant effective
action~\cite{Le85}. However, because both set of currents correspond
to same formal definition, they must coincide in their ultraviolet
convergent terms, and thus they can only differ by a local polynomial
of dimension at most $D-1$~\cite{Ba84,Sa96}
\begin{equation}
\bfJ_i(x)=\bfJ_i^c(x)+\bfQ_i(x)\,.
\end{equation}
Here, $\bfJ_i$ is consistent, $\bfJ_i^c$ is covariant and $\bfQ_i$ is
a polynomial. Note that the arguments used to reach this relation hold
in particular for the class of non local Dirac operators considered in
this work. This relation is already sufficient to show that the
essential anomaly only contains vector and axial
fields~\cite{Sa96}. Indeed, let us separate the Dirac operator $\bfD$
into two components $\bfD_0$ and $\bfN$ both transforming covariantly
under chiral transformations and such that $\bfD_0$ contains the term
$\gamma_\mu P_\mu$. Consequently, the vector and axial fields should
also be in $\bfD_0$ since they mix with $\gamma_\mu P_\mu$ under
chiral transformations. Also note that $\bfD_0$ is a valid Dirac
operator whereas $\bfN$ by itself is not. The change in the effective
action due to passing from $\bfD_0$ to $\bfD=\bfD_0+\bfN$ can be
obtained by integrating the consistent current along the path
$\bfD_t=\bfD_0+t\bfN$ with $0\le t \le 1$. The result does not depend on
the particular interpolating path since integrability conditions are
satisfied. The contribution from the
covariant current will be invariant and does not change the
anomaly. The contribution from the polynomial current will be a
polynomial. This latter term is responsible for the change in the
anomaly introduced by adding $\bfN$. Since this change derives from a
polynomial it is removable by counter terms. Therefore taking
$\bfD_0=\gamma_\mu(P_\mu+V_\mu+A_\mu\gamma_5)$ yields an anomaly
involving only vector and axial fields. This anomaly will be minimal
by using a chiral covariant renormalization for the real part of the
effective action, such as the $\zeta$-function prescription applied to
$-\frac{1}{2}\tr\log(\bfD\bfD^\dagger)$.

The actual construction of the counter terms can be done using the
method in ref.~\cite{Sa96}. In order to keep the reasoning straight we
will skip some technicalities and use a rather symbolic notation. The
current can be written as $\bfJ=\partial W/\partial\bfD$. Under
vector and axial
transformations, the covariant current transforms as
$\bfJ^c\to e^{-i\beta}\bfJ^c e^{i\beta}$ and
$\bfJ^c\to e^{i\alpha\gamma_5}\bfJ^c e^{i\alpha\gamma_5}$
 respectively so that $\langle
\bfN\bfJ^c\rangle$ remains invariant in both cases. (Recall that $\bfN$
transforms as $e^{i\beta}\bfN e^{-i\beta}$ and $e^{-i\alpha\gamma_5}\bfN
e^{-i\alpha\gamma_5}$ respectively.) Infinitesimally this implies
\begin{eqnarray}
0=\bar\delta_V\bfJ^c &:=& \delta_V\bfJ^c-[i\beta,\bfJ^c ]\,,
\\
0=\bar\delta_A\bfJ^c &:=& \delta_A\bfJ^c-\{i\alpha\gamma_5,\bfJ^c\}\,,
\end{eqnarray}
where we have introduced the covariant vector and axial variations
$\bar\delta_V$ and $\bar\delta_A$ respectively.
On the other hand, since the polynomial current $\bfQ$ accounts for
changes in the anomaly induced by changes in $\bfD$, it should satisfy
the following two equations
\begin{equation}
\bar\delta_V\bfQ = 0 \,, \qquad
\bar\delta_A\bfQ = \frac{\partial{\cal A}_A}{\partial\bfD} \,.
\label{eq:38}
\end{equation}
Once these equations are solved, the counter terms needed to remove the
extra contribution coming from $\bfN$ are
\begin{equation}
W_{\rm ct}= (4\pi)^{D/2}\int_0^1 dt\langle\bfN\bfQ_t\rangle\,,
\end{equation}
where $\bfQ_t$ is the polynomial current corresponding to
$\bfD_t=\bfD_0+t\bfN$. (Recall that we are using a schematic
notation. In an actual calculation one has to distinguish each of the
external fields which define $\bfD$ and their associated currents as
discussed below.)

The right hand side of the second eq.~(\ref{eq:38}) can be computed
from the known anomaly, eqs.~(\ref{eq:26},\ref{eq:27}). For instance,
in two dimensions
\begin{equation}
\frac{\partial{\cal A}_A}{\partial\bfD_L}=
\frac{1}{4\pi}\{2i\alpha\gamma_5,\bfD_L\}\,.
\label{eq:39}
\end{equation}
(The cyclic property has been used since it turns out to be justified
in this case.) In eq.~(\ref{eq:38}) $\bfQ$ is the unknown. If the
anomaly were the variation of a polynomial action, an immediate
solution would be given by the corresponding polynomial
current. Because the anomaly contains an essential part, such
polynomial action does not exist. Remarkably, at a formal level there
is a polynomial action from which the axial anomaly
derives~\cite{Sa96}, namely,
\begin{equation}
W_0 = -\langle{1 \over 2}\bfD_L^2\rangle
\end{equation}
in two dimensions and
\begin{eqnarray}
W_0 &=& -\langle {1\over 24}\bfD_L^4 +{1\over
24}\bfD_L^2\gamma_\mu\bfD_L^2\gamma_\mu + {1\over
12}\bfD_L^3\gamma_\mu\bfD_L\gamma_\mu
\\ &&
+\frac{1}{8}(
 \gamma_\mu\bfMt_\mu\bfD_L^2
+\gamma_\mu\bfD_L^2\bfMt_\mu
+\gamma_\mu\bfD_L\bfMt_\mu\bfD_L
+\bfMt_\mu^2
+\{\bfD_L,\bfMt_{\mu\mu}\}
)\rangle
\end{eqnarray}
in four dimensions. In this context ``formal level'' means that the
correct anomaly is obtained from $W_0$ if the cyclic property is
used. Actually, the operators involved in $W_0$ are not multiplicative
and so the cyclic property does not hold (it does hold in Dirac and
flavor spaces but not for differential operators in coordinate space).
Even so, the action
$W_0$ has a unique well-defined current $\bfQ_0=\partial
W_0/\partial\bfD$, which is obtained from (using the cyclic property
to put all $\delta\bfD$ together)
\begin{equation}
\delta W_0= (4\pi)^{D/2}\langle\delta\bfD\,\bfQ_0\rangle\,.
\end{equation}
$\bfQ_0$ is well-defined in the sense that it is unchanged under
cyclic permutations of the operators in $W_0$. For instance, in two
dimensions, $\delta W_0= -\langle\delta\bfD_L\,\bfD_L\rangle$ and so
\begin{equation}
\bfQ_0=-\frac{1}{4\pi}\bfD_L\,.
\label{eq:45}
\end{equation}
Because $W_0$ formally gives the anomaly, $\bfQ_0$ is a solution of
eq.~(\ref{eq:38}). For instance, in two dimensions, $\bar\delta_A
\bfQ_0$ is easily computed and gives precisely the right hand side of
eq.~(\ref{eq:39}). This solution is nevertheless formal because
$\bfQ_0$ is not a multiplicative operator in general and $\bfQ$ must
be multiplicative. This must be solved by subtracting to $\bfQ_0$ a
polynomial current $\bfQ_1$ which transforms covariantly (i.e.,
$\bar\delta_A \bfQ_1=0$) and such that $\bfQ=\bfQ_0-\bfQ_1$ is
multiplicative. Again, using the two dimensional case as an example,
one can consider a new local Dirac operator
$\bar{\bfD}_L=-\bfD_L^\dagger$. The minus sign ensures that it is an
admissible Dirac operator, i.e., there is a corresponding
$\bar{\bfY}$. The adjoint implies that it transforms as a covariant
current under axial transformations. Then it can immediately be
checked that the covariant polynomial current
\begin{equation}
\bfQ_1=-\frac{1}{4\pi}\bar{\bfD}_L\,,
\end{equation}
has the same non multiplicative part as $\bfQ_0$ in eq.~(\ref{eq:45}).
(In this case this is trivial to see. As noted above, the best way to
make this check in general is to replace $P_\mu$ by $P_\mu+a_\mu$ and
see that $\bfQ_0-\bfQ_1$ is $a_\mu$-independent.)

To consider the non local case in detail, we introduce the currents as
\begin{equation}
\delta W=(4\pi)^{D/2}\langle
\delta\bfY\bfJ_L
+\delta\bfM_\mu\bfJ_\mu
+\frac{1}{2}\delta\bfM_{\mu\nu}\bfJ_{\mu\nu}
+\cdots\rangle\,.
\label{eq:47}
\end{equation}
In four dimensions, the calculation proceeds by computing the non
multiplicative polynomial formal currents
\begin{equation}
\delta W_0=(4\pi)^2\langle
\delta\bfY\bfQ^0_L
+\delta\bfM_\mu\bfQ^0_\mu
+\frac{1}{2}\delta\bfM_{\mu\nu}\bfQ^0_{\mu\nu}
\rangle \,.
\end{equation}
The polynomials $\bfQ_0$ are well defined and have the correct axial
transformation but they are non multiplicative. In order to construct
the corresponding $\bfQ_1$, let us introduce the Dirac operator
\begin{equation}
\bar{\bfD}=-\bfD^\dagger\,.
\end{equation}
Such operator belongs to the same class as $\bfD$, that is, it can be put
as $\bar{\bfD}_L+\bar{\bfM}$ where $\bar{\bfM}$ has the form given in
eq.~(\ref{eq:12}). The corresponding coefficients are easily computed
by noting that
\begin{equation}
\tilde{\bar{\bfM}}(p)=e^{ipX}\bar{\bfM}e^{-ipX}=-(\tilde{\bfM}(p))^\dagger\,.
\end{equation}
And so, $\tilde{\bar{\bfM}}_{\mu_1\dots\mu_n}=
-\bfMt^\dagger_{\mu_1\dots\mu_n}$. For the two lowest
coefficients one finds
\begin{eqnarray}
\bar{\bfM}_\mu &=& -\bfM^\dagger_\mu \,, \nonumber \\
\bar{\bfM}_{\mu\nu} &=& -\bfM^\dagger_{\mu\nu}
-t_{\mu\nu\rho\sigma}[P_\rho,\bfM^\dagger_\sigma]\,.
\end{eqnarray}
The form of $\bar{\bfD}$ is such that axially it transforms as a
covariant current. The currents $\bfQ_1$ are defined by suitably
replacing some of the $\bfD$ in $\bfQ_0$ by $\bar{\bfD}$, so that
$\bfQ_1$ transforms covariantly. For instance, terms of the form
$\bfMt_{\mu\mu}$, $\gamma_\mu\bfMt_\mu\bfD_L$, $P_\mu\bfD_L$ or
$\gamma_\mu\bfD_L^2$ in $\bfQ_0$ correspond, respectively, to
$\tilde{\bar{\bfM}}_{\mu\mu}$, $\gamma_\mu\bfMt_\mu\bar{\bfD}_L$,
$P_\mu\bar{\bfD}_L$ and $\gamma_\mu\bfD_L\bar{\bfD}_L$ in $\bfQ_1$. It
is an exercise to check that $\bfQ=\bfQ_0-\bfQ_1$ is multiplicative.

There is a technical subtlety in checking that $\bfQ_1$ is actually an
axial covariant current. This is because the coefficients
$\bfM_{\mu_1\dots\mu_n}$ are multiplicative but not axially
covariant. Therefore, the covariance of a current does not directly
correspond to the equation $\bar\delta_A\bfJ^c_{\mu_1\dots\mu_n}=0$.
In order to derive the correct equations, the simplest method is to
introduce the currents $\bfJt$ corresponding to the covariant
quantities $\bfMt$,
\begin{equation}
\delta W=(4\pi)^{D/2}\langle
\delta\bfY\bfJ_L
+\delta\bfMt_\mu\bfJt_\mu
+\frac{1}{2}\delta\bfMt_{\mu\nu}\bfJt_{\mu\nu}
+\cdots\rangle\,.
\end{equation}
They can be related to the currents $\bfJ$ by comparing with
eq.~(\ref{eq:47}). For the two lowest orders
\begin{eqnarray}
\bfJt_\mu &=&
\bfJ_\mu-\frac{1}{2}t_{\mu\nu\rho\sigma}P_\nu\bfJ_{\rho\sigma}+\cdots \,,
\nonumber \\
\bfJt_{\mu\nu} &=& \bfJ_{\mu\nu}+\cdots \,.
\end{eqnarray}
Note that the currents $\bfJt$ are covariant (up to anomaly) but not
multiplicative due to the presence of $P_\mu$. The covariant part
satisfies $\bar\delta_A\bfJt^c=0$.  For the polynomial part in four
dimensions, the relations become
\begin{eqnarray}
\bfQt_\mu &=&
\bfQ_\mu-\frac{1}{2}t_{\mu\nu\rho\sigma}P_\nu\bfQ_{\rho\sigma} \,,
\nonumber \\
\bfQt_{\mu\nu} &=& \bfQ_{\mu\nu}\,.
\end{eqnarray}
A straightforward calculation shows that $\bar\delta_A\bfQt_1=0$ and
so the $\bfQ_1$ constructed above have the correct axial
transformation. This completes the construction of the counter terms in
the non local case.

\section{Trace anomaly}\label{sec:4}

The scale transformation $\psi(x)\to
e^{-\alpha_S(D-1)/2}\psi(e^{-\alpha_S}x)$ can be compensated by a
corresponding transformation in $\bfD$, namely,
\begin{equation}
\bfY(x)\to e^{-\alpha_S}\bfY(e^{-\alpha_S}x)\,,\quad
\bfM_{\mu_1\dots\mu_n}(x) \to
e^{-\alpha_S(n+1)}\bfM_{\mu_1\dots\mu_n}(e^{-\alpha_S}x)\,.
\end{equation}
Therefore, scale transformations are a symmetry of our class of non
local Dirac operators. The linear operator which produces the scale
transformation $\psi\to \Omega_S\psi$ can be written as
$\Omega_S=e^{-\alpha_S((D-1)/2+X\partial)}$, and thus the Dirac
operator transforms as $\bfD\to\Omega_S^{-1\dagger}\bfD\Omega_S^{-1}$.
Infinitesimally it implies
\begin{equation}
\delta_S\bfD= -\alpha_S(\bfD-i[X_\mu P_\mu,\bfD])\,.
\end{equation}
The corresponding trace anomaly, within the $\zeta$-function method
is~\cite{Sa96}
\begin{equation}
{\cal A}_S= \delta_S W= \alpha_S\Tr(\bfD^s)_{s=0}\,.
\end{equation}
The calculation is entirely similar to that of the axial anomaly. In
two dimensions one finds
\begin{equation}
{\cal A}_S= \alpha_S\langle
\bfD_L^2+\frac{1}{4}\{\gamma_\mu,\bfD_L\}^2
+\gamma_\mu\bfM_\mu
\rangle\,,
\end{equation}
and in four dimensions
\begin{eqnarray}
{\cal A}_S &=& \alpha_S\Big\langle
{1\over 2}\bfD_L^4 +
{1\over 12}(\bfD_L^2\{\gamma_\mu,\bfD_L\}^2+
\{\gamma_\mu,\bfD_L\}\bfD_L^2\{\gamma_\mu,\bfD_L\}
+\{\gamma_\mu,\bfD_L\}^2\bfD_L^2)
  \nonumber\\ &&
+ {1\over 96} (\{\gamma_\mu,\bfD_L\}^2\{\gamma_\nu,\bfD_L\}^2
+ (\{\gamma_\mu,\bfD_L\}\{\gamma_\nu,\bfD_L\})^2
+ \{\gamma_\mu,\bfD_L\}\{\gamma_\nu,\bfD_L\}^2\{\gamma_\mu,\bfD_L\})
  \nonumber\\ &&
+\frac{1}{12}(
 \gamma_\mu\bfMt_\mu\bfD_L^2
+\gamma_\mu\bfD_L^2\bfMt_\mu
+\gamma_\mu\bfD_L\bfMt_\mu\bfD_L
)
  \nonumber\\ &&
-\frac{1}{24}(
 \bfMt_\mu\gamma_\mu\bfD_L^2
+\bfD_L^2\gamma_\mu\bfMt_\mu
+\bfD_L\{\gamma_\mu,\bfMt_\mu\}\bfD_L
+\bfD_L\gamma_\mu\bfD_L\bfMt_\mu
+\bfMt_\mu\bfD_L\gamma_\mu\bfD_L
)
  \nonumber\\ &&
+\delta_{\mu\nu\alpha\beta}\Big(
\frac{1}{36}(
 \gamma_\mu\bfMt_\nu\gamma_\alpha\bfD_L\gamma_\beta\bfD_L
+\gamma_\mu\bfD_L\gamma_\alpha\bfMt_\nu\gamma_\beta\bfD_L
+\gamma_\mu\bfD_L\gamma_\alpha\bfD_L\gamma_\beta\bfMt_\nu
)
  \nonumber\\ &&
+\frac{1}{24}\gamma_\mu\bfMt_\nu\gamma_\alpha\bfMt_\beta
+\frac{1}{24}\{\gamma_\alpha\bfMt_{\mu\nu},\gamma_\beta\bfD_L\}
+\frac{1}{12}\gamma_\mu\bfMt_{\nu\alpha\beta}
\Big)
\Big\rangle\,.
\end{eqnarray}
Where $\delta_{\mu\nu\alpha\beta}=
\delta_{\mu\nu}\delta_{\alpha\beta}+\delta_{\mu\alpha}\delta_{\nu\beta}
+\delta_{\mu\beta}\delta_{\alpha\nu}$. The result is again a local
polynomial of dimension $D$ in the external fields and their
derivatives. Unlike the axial case, the coefficients $\bfM_\mu$ in two
dimensions and $\bfM_{\mu\nu\alpha}$ in four dimensions do contribute
to the scale anomaly.

Because scale and chiral transformations commute (in a properly defined
sense), the crossed variations $\delta_S{\cal A}_{V,A}$ and
$\delta_{V,A}{\cal A}_S$ coincide and they vanish since the axial
anomaly is scale invariant. Thus the scale anomaly must be chiral
invariant. The vector gauge invariance of the previous expressions is
easy to check noting that the operators inside $\langle\,\rangle$ are
multiplicative. Axial invariance is much more involved in general. In
the two dimensional case, it is immediate to see that
$\langle\gamma_\mu\bfM_\mu\rangle$ is invariant. In four dimensions it
is relatively easy to check that the trace anomaly is axially
invariant in the particular case of $\bfM_\mu=0$, which, as noted
previously, defines a class of operators invariant under chiral and
scale transformations.

The scale anomaly is already minimal. It can be modified by adding
polynomial counter terms of dimension smaller than $D$ but this would
add terms of the same type to the scale anomaly.

\section*{Acknowledgments}
This work is supported in part by funds provided by the Spanish DGICYT
grant no. PB95-1204 and Junta de Andaluc\'{\i}a grant no. FQM0225.

\end{document}